\documentclass[aps,pra,showpacs,twoside,twocolumn,longbibliography,10pt]{revtex4-1}
\usepackage[colorlinks=true, citecolor=red, urlcolor=blue ]{hyperref}
\usepackage{epsfig,newlfont,amssymb,amsfonts,amsmath,bm,subfigure,palatino,mathtools,amsthm,braket,times,soul,enumitem,color}
\usepackage[normalem]{ulem}
\newcommand{\stkout}[1]{\ifmmode\text{\sout{\ensuremath{#1}}}\else\sout{#1}\fi}
\usepackage[english]{babel}
\usepackage[utf8]{inputenc}
\usepackage{array}
\usepackage{xcolor}
\usepackage{graphics}

\newcommand{\ketbra}[2]{|#1\rangle \langle #2|}
\def\Tr{\text{Tr}}

\usepackage{amsthm}
\usepackage{verbatim}
\usepackage{bbm}
\usepackage{wrapfig}
\usepackage{makecell}
\usepackage{tikz}
\usetikzlibrary{arrows.meta, decorations.pathreplacing, calc, shapes.geometric}

\usepackage{orcidlink}

\usepackage{hyphenat}
    \usepackage{amsmath}

\newlength\figureheight 
\newlength\figurewidth

\begin{document}
\title{Scaling vs entanglement in measurement-induced phase transition for non-integrable systems}

\author{
Paranjoy Chaki$^{1,2}$, 
Protyush Nandi$^{1,3}$, 
Subinay Dasgupta$^{1}$, 
Ujjwal Sen$^{1,2}$, 
}

\affiliation{
$^{1}$Harish-Chandra Research Institute,  
Chhatnag Road, Jhunsi, Prayagraj 211019, India\\
\(^2\) Homi Bhabha National Institute,  Training School Complex, Anushakti Nagar, Mumbai 400 094, India\\
$^{3}$University of Calcutta, 92 Acharya Prafulla Chandra Road,
Kolkata 700009, India
}


\begin{abstract}
We find that  the measurement-induced phase transition generated by deterministic global measurements, previously observed in the integrable transverse-field Ising model (TFIM), persists in non-integrable variants of the same. To address this question, we consider the TFIM with longitudinal field and the axial next-nearest-neighbor Ising (ANNNI) model. We show that both the survival probability and the bipartite entanglement entropy consistently capture a transition between area-law and volume-law entangled phases for two distinct initial states: a product state with all spins polarized along the transverse direction
and a Greenberger–Horne–Zeilinger (GHZ) state. Finite-size scaling reveals a pronounced initial-state dependence: for the polarized product state, the transition point follows an inverse-square-root scaling with system size in both non-integrable models, consistent with the integrable TFIM, whereas for the GHZ initial state, it deviates from this scaling and approaches zero considerably more slowly in the non-integrable models than in the integrable TFIM. These results establish the robustness of measurement-induced transitions under deterministic measurements against integrability breaking while highlighting the crucial role of the initial state in governing their scaling behavior.
\end{abstract}

\maketitle
\section{Introduction}

Measurement-induced phase transitions (MIPTs) constitute a fascinating class of nonequilibrium phenomena that has been extensively investigated across a wide range of platforms, including random quantum circuits and quantum spin chains subjected to diverse measurement protocols.

To the best of our knowledge, the measurement-induced phase transition was first introduced in Ref.~\cite{SDG_2}, where its existence was demonstrated through the behavior of the survival probability. Subsequently, this phenomenon was investigated in random quantum circuits composed of random two-qubit unitary gates interspersed with local projective measurements performed probabilistically~\cite{zeno_1,Skinner,MIPT_2}. The competition between the disentangling action of local measurements and the entanglement-generating dynamics of random two-qubit unitary gates gives rise to a transition between area-law and volume-law entanglement phases. Consequently, MIPTs have attracted considerable attention across a wide variety of measurement protocols and random-circuit architectures~\cite{MIPT_2,MIPT_3,Chan2019,Szyniszewski2019,Bao2020,Choi2020,Gullans2020,GullansHuse2020,Jian2020,Zabalo2020,Iaconis2020,Turkeshi2020,Zhang2020,SzyniszewskiII2020,Nahum2021,Ippoliti2021,IppolitiII2021,lavasani2021,LavasaniII2021,Sang2021,Block2022,Sharma2022,Agrawal2022,Baratt2022,jian2023,Shane2023}. The investigation of MIPTs is, however, not restricted to random-circuit models; these transitions have also been explored in transverse-field Ising models~\cite{Isi_1,Isi_2,Weak_Ising,Su2024,Tirrito2023,Roser2023,Biella2021,Tista2024,Pavigilianti2023}, trapped-ion systems~\cite{exp_1}, many-body-localized systems~\cite{MIPT_MBL}, fermionic systems~\cite{Buchhold2021,Poboiko2024,Muller2022,Chatterjee2024,Minato2022,Jin2024,Kaijian2021,FT}, higher-dimensional systems~\cite{Turkeshi2020,geo_2,Nahum2021,geo_4,geo_5}, and superconducting-qubit platforms~\cite{koh2023,exp_4}. Furthermore, beyond entanglement-based diagnostics, MIPTs can also be identified through the behavior of the survival probability~\cite{SDG_2,Suv_sp}.
 In Ref.~\cite{Suv_sp}, we considered a transverse field Ising chain with nearest-neighbor interactions along the $X$ direction and a transverse magnetic field along the $Z$ direction. The system is initially prepared with all spins polarized along the transverse ($+Z$) direction and subsequently evolves under the transverse-field Ising Hamiltonian (TFIM) for a time interval $\tau$. This unitary evolution is followed by a global measurement onto the complementary subspace orthogonal to the initial configuration. We demonstrated that the survival probability, defined here as the probability of obtaining a component other than the initial configuration, captures a convex-to-concave transition as a function of the measurement interval. Remarkably, the transition is also captured by bipartite entanglement, revealing that the transition is not only a convex-to-concave transition but also an area-to-volume law transition.


In the present work, we investigate whether the measurement-induced transition persists beyond the integrable TFIM when the system's evolution is governed by non-integrable Hamiltonians and also the role of an initially entangled state rather than restricting our analysis only to $+Z$-polarized product state. To address this question, we consider two non-integrable models: the transverse-field Ising model supplemented by a longitudinal field along the $X$ direction~\cite{mix_field} and the axial next-nearest-neighbor Ising (ANNNI) model~\cite{sdg_ANNNI}. The spin system is initially prepared in a generalized GHZ state, which can be continuously tuned between the $+Z$-polarized product state at one limiting point and the maximally entangled GHZ state at the other. We demonstrate that, for both non-integrable models, the survival probability exhibits a convex-to-concave transition at a characteristic critical measurement period for both the $+Z$-polarized and maximally entangled GHZ initial states. It implies that the MIPT protocol remains robust under the non-integrable Hamiltonians for both types of considered initial configurations.

Furthermore, we show that the area-to-volume law entanglement transition occurs approximately at the same critical measurement rate at which the convex-to-concave transition in the survival probability is observed. This correspondence holds for both initial states and for both non-integrable models considered. For completeness and comparison, we also present the corresponding analysis for a GHZ initial state evolving under the integrable TFIM without a longitudinal field, which is not studied previously.

On the other hand, for initially polarized spin configuration in the transverse ($+Z$) direction, we perform the scaling analysis of the transition point, obtained from the survival probability, up to the system sizes $L=1000$ and show that a qualitatively similar scaling as the integrable TFIM. We also show that, although the transition point scales inversely proportional to the square root of system size, the proportionality constant decreases with the increase of the strength of the non-integrable term, implying that the transition point goes to zero faster for non-integrable models than for TFIM when initially the state is an up-polarized state.

On the other hand, when the initial state is the GHZ state, the main distinction we highlight is that, for all three models, the scaling of the transition point with system size does not exhibit the inverse-square-root dependence on the system size. Moreover we show that although the transition point moves toward zero with system size for all three models, the movement towards zero is much slower for the non-integrable models than the integrable TFIM. It implies that in non-integrable models the transition persists for larger system sizes than the integrable TFIM. 

The remainder of the paper is organized as follows. In Sec.~\ref{prot}, we introduce the measurement protocol employed throughout this work. Section~\ref{up_in_z} presents the results for the fully $Z$-polarized initial state, with Secs.~\ref{IIIA} and~\ref{IIIB} devoted to the survival-probability and entanglement-entropy analysis, respectively. In Sec.~\ref{IV}, we extend our investigation to the GHZ initial state; the corresponding survival-probability and entanglement-entropy results are discussed in Secs.~\ref{IVA} and~\ref{IVB}, respectively. The finite-size scaling analysis for the fully $Z$-polarized and GHZ initial states are presented in Secs.~\ref{VA} and~\ref{VB}, respectively. Finally, we summarize our main findings and conclude in Sec.~\ref{VI}.

\section{protocol}\label{prot}

In this section, we discuss our measurement based protocol. In this work we have considered two non-integrable Hamiltonians, which generate the time evolution unitary $\exp(-i\mathcal{H}t)$  of our protocol. The Hamiltonians are

\begin{equation}
\mathcal{H} = \sum_i \sigma_i^x \sigma_{i+1}^x
- h_z \sum_i \sigma_i^z
- h_x \sum_i \sigma_i^x .
\label{eq:TFIM}
\end{equation}

\begin{equation}
\mathcal{H} = - \sum_i \sigma_i^x \sigma_{i+1}^x
+ \kappa \sum_i \sigma_i^x \sigma_{i+2}^x
- h_z \sum_i \sigma_i^z .
\label{eq:ANNNII}
\end{equation}

\begin{figure}
		\centering
			\includegraphics[width=8.5cm]{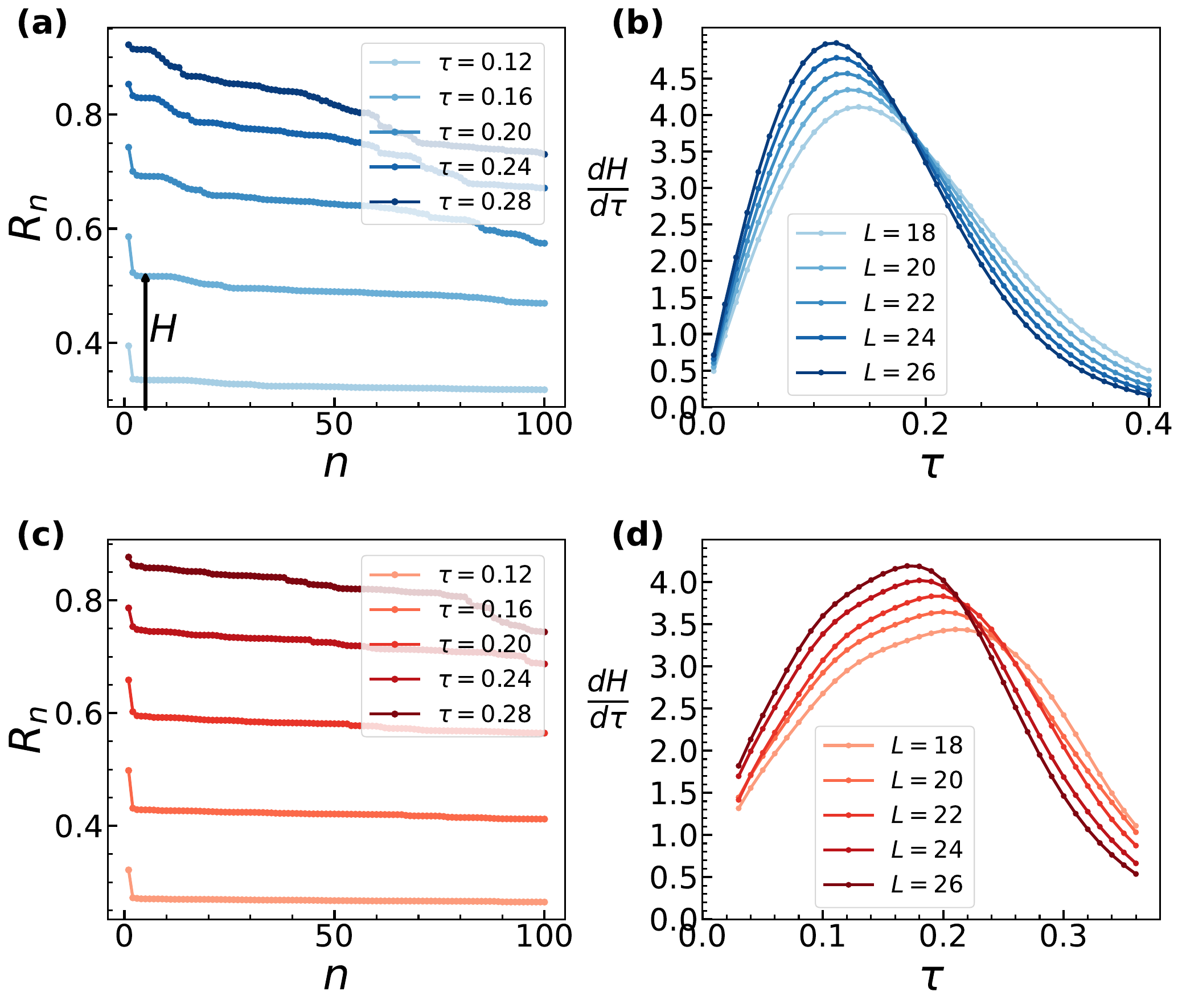}
\caption{\textbf{Behaviour of the survival probability with the number of measurements.}
Panels (a) and (c) show the survival probability ($R_n$) as a function of the number of measurements ($n$) for $\tau=0.12, 0.16, 0.20, 0.24,$ and $0.28$ with system size $L=26$, corresponding to the transverse-field Ising model with a longitudinal field and the ANNNI model, respectively where $\tau$ is the period of measurement.  For the Ising model, the parameters are chosen as $h_z=0.5$ and $h_x=0.6$, while for the ANNNI model the parameters are $\kappa=0.2$ and $h_z=0.2$. The different values of $\tau$ are represented by sequential blue and red color gradients in panels (a) and (c) respectively, as indicated in the legends. Panels (b) and (d) display $\frac{dH}{d\tau}$ as a function of $\tau$ at time step $n=3$ for the transverse-field Ising model with longitudinal field and the ANNNI model, respectively, for system sizes $L=18, 20, 22, 24,$ and $26$. The curves in panels (b) and (d) are shown using sequential blue and red color schemes, respectively. The Hamiltonian parameters used in panels (b) and (d) are the same as those employed in panels (a) and (c). } 
		\label{fig11}
		\end{figure}

The models in Eq.~\eqref{eq:TFIM} and Eq.~\eqref{eq:ANNNII} describe the transverse field Ising model (TFIM) with longitudinal field ($h_x$) and the axial next nearest neighbour Ising (ANNNI) model respectively. The longitudinal field and the next-to-next neighbor interaction term in Eq.~\eqref{eq:TFIM} and Eq.~\eqref{eq:ANNNII} are responsible for integrability breaking.  $h_x$ and $\kappa$ are the parameters that introduce non-integrability in the integrable TFIM in Eq.~\eqref{eq:TFIM} and Eq.~\eqref{eq:ANNNII} respectively.



Starting from some initial state $\ket{\psi_{in}}$, we allow the system to evolve for a period of time $\tau$. Then perform a global measurement in the basis $\{\ketbra{\psi}{\psi}_{in}, \mathbbm{I}-\ketbra{\psi}{\psi}_{in}\}$ where the action of the measurement operator $\mathbbm{I}-\ketbra{\psi}{\psi}_{in}$ takes the initial state $\ket{\psi_{in}}$ to it's orthogonal complementary space. Here we took the outcome corresponding to the measurement operator $\mathbbm{I}-\ketbra{\psi}{\psi}_{in}$. The process of unitary evolution and projection global measurement in the orthogonal complementary space is repeated throughout the dynamics.

We will use two quantities, survival probability and entanglement entropy, to analyze the dynamics described above. Let us consider $\ket{\psi_{1}},\ket{\psi_{2}},\ket{\psi_{3}},...,\ket{\psi_{n}} $ are the post-selected states up to the $n$th measurement. Then the $n$th post-selected state can be expressed as
 
\begin{equation}
|\psi_n\rangle
=
e^{-i\mathcal{H}\tau} |\psi_{n-1}\rangle
-
\langle \psi_{in} | e^{-i\mathcal{H}\tau} | \psi_{n-1} \rangle
\, |\psi_{in}\rangle .
\end{equation}

The measurement process can be thought of as the question, ``Is the state in the initial state?'' We are following that trajectory where the answer is always ``No''. The survival probability is the probability of getting ``No'' after each measurement. 
Mathematically, it can be expressed as 
\begin{equation}
R_n=\braket{\psi_n |\psi_n}.
\end{equation}
In our work, we consider the von Neumann entropy as a reliable measure of entanglement. Let us consider a many-body quantum state $\ket{\psi_n}$ that consists of $L$ number of sites. We divide the system into two bipartitions; one contains $l$ numbers of sites, and the rest contain $L-l$ numbers of sites. Then the Von-Neumann or entanglement entropy between two parties is given by
        \begin{equation}
            S=-\Tr [\rho_l\ln\rho_l]
        \end{equation}
where $\rho_l$ is the reduced density matrix of the segment of length $l$.\\
By using the two quantities $R_n$ and $S$, we will show the persistence of the measurement-induced transition, present in the integrable TFIM \cite{Suv_sp}, even in non-integrable models. We will study how the transition point scales with the system size and how the scaling changes with the presence of entanglement in the initial state. \\

In this work, we shall consider two types of initial state. One is $\ket{\psi_{in}}=\ket{0}^{\otimes L}$, where all spins are along Z direction. This state was used in our previous study \cite{Suv_sp}. The other one is 
the generalized GHZ state 
\begin{equation}\label{eq6}
|\psi_0\rangle^{\mathcal{G}}_{GHZ} = \cos \phi\, |U\rangle + \sin \phi\, |D\rangle,
\end{equation}
where $|U\rangle := \ket{0}^{\otimes L}, 
\hspace{0.2cm}$ and 
$|D\rangle := \ket{1}^{\otimes L}$. Here $\phi \in[0,\pi/2]$ is a parameter. For $\phi=\pi/4$ the state become maximally entangled GHZ state.

\section{Detection of transition points for the fully transverse-polarized initial state}\label{up_in_z}

In this section, we consider the initial configuration such that all the spins are pointed in the $+Z$ (transverse) direction,~i.e., $\ket{\psi_{in}}=\ket{0}^{\otimes L}$. We consider both survival probability and entanglement entropy to detect the transition point.

\subsection{Detection of transition point by survival probability for transverse-polarized initial state}\label{IIIA}

In this subsection, we describe the behavior of survival probability under periodic deterministic measurement for both the mixed-field TFIM and the ANNNI model. In panel (a) of Fig.~\ref{fig11}, we plot $R_n$ vs. $n$ for the TFIM with longitudinal field for different $\tau$ values, viz., $\tau=0.12,0.16,0.20,0.24,0.28$. The curves corresponding to different $\tau$ values are denoted by sequential blue colors. We consider the value of transverse and longitudinal fields as $h_z=0.5$ and $h_x=0.6$, respectively. We found that the $R_n$ versus $n$ curve exhibits at least one interval, $n_1<n<n_2$, over which $R_n$ remains approximately constant, forming a plateau. The corresponding value of $R_n$ in this region is defined as the plateau height ($H$) as denoted in the panel (a) of Fig.~\ref{fig11} corresponding to the curves having the $\tau$ value $\tau=0.16$ by black arrow. The figure shows a clear increment of the height of the plateau, and to observe the behavior of the height increment with $\tau$, we plot $\frac{dH}{d\tau}$ vs. $\tau$ graph in panel (b) for system sizes $L=18,20,22,24,26$, and the plot shows a peak, which indicates a clear convex-to-concave transition of the height of the plateau with $\tau$, around the transition point $\tau_c\approx0.13$. We have done the same analysis for the ANNNI model considering the parameters $\kappa=0.2$ and $h_z=0.2$ in panels (c) and (d). The curves in panels (c) and (d) for different parameters are shown by sequential red colors. 
We found a similar feature as that obtained for the TFIM with  longitudinal field. A clear convex-to-concave transition is captured in the panel (d) around the $\tau$ value $\tau_c\approx 0.17$.

Our analysis indicates that the convex-to-concave transition in the height of survival probability curve is not only captured in the integrable Ising model but also present when integrability is broken by both longitudinal field in the mixed-field TFIM and next-to-next-nearest neighbor interaction term in the ANNNI model.

\subsection{Detection of transition point by entanglement corresponding to transverse-polarized initial state}\label{IIIB}
\begin{figure}
		\centering
			\includegraphics[width=8.5cm]{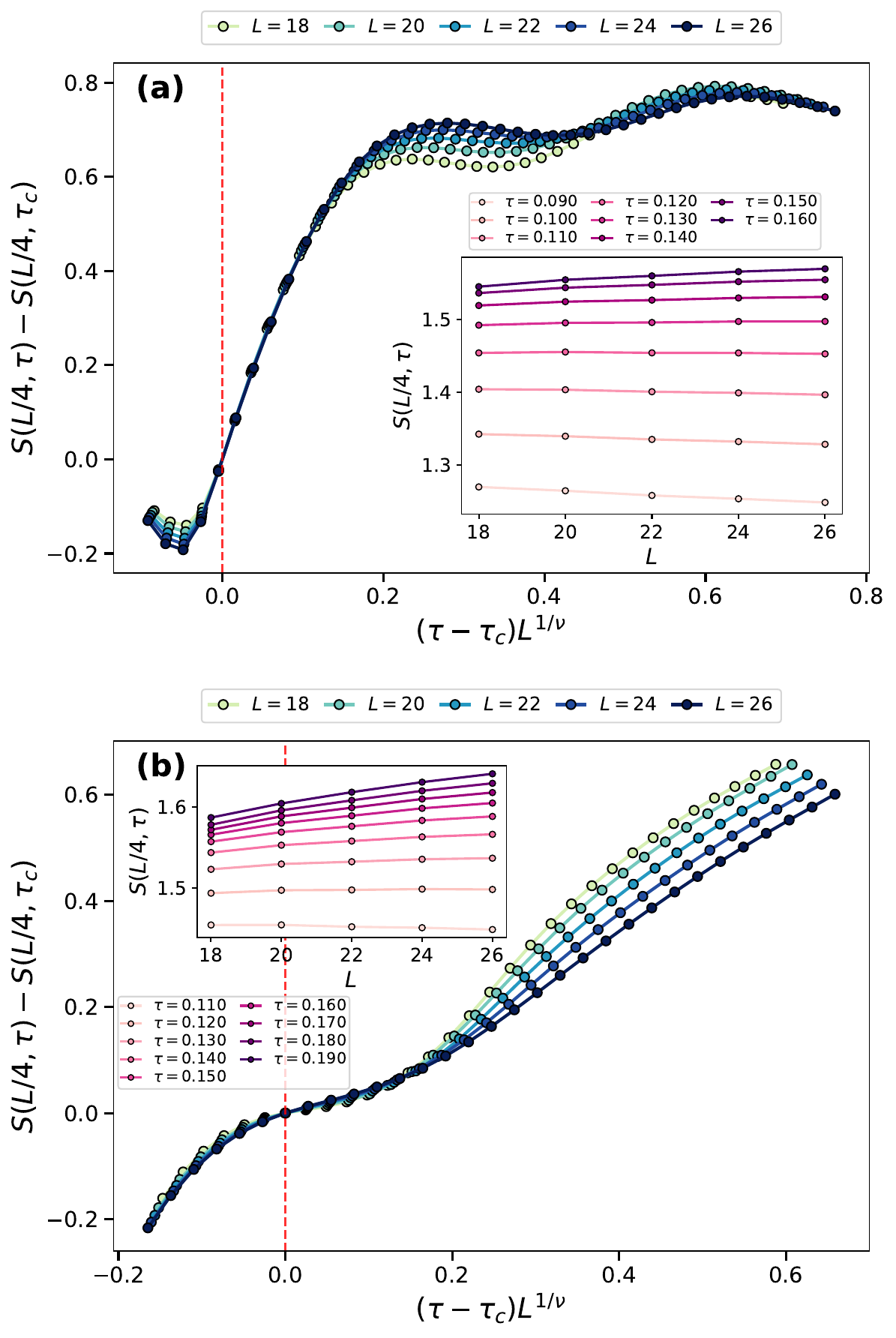}
\caption{\textbf{Behavior of entanglement with system size for initially $+Z$ polarized initial state.}
Panel (a) shows the scaling collapse of entanglement entropy for transverse field Ising model with longitudinal field with $h_z=0.5$ and $h_x=0.6$. To that end, we consider the horizontal axis as $(\tau-\tau_c)L^{1/\nu}$ and the vertical axis is $S(L/2,\tau)-S(L/2,\tau_c)$. Here we have considered the system sizes $L=18,20,22,24,26$. In the inset, we plot $S(L/4,\tau)$ vs $L$ for different $\tau$ values given by $\tau=0.90,0.10,0.11,...,0.14$ respectively. Panel (b) corresponds to the $S(L/2,\tau)-S(L/2,\tau_c)$ vs. $(\tau-\tau_c)L^{1/\nu}$ graph for the ANNNI model for the parameters $\kappa=0.2$ and $h_z=0.2$ for system sizes $L=18,20,22,24,26$.  In the inset we plot $S(L/4,\tau)$ vs $L$ for $\tau$ values $\tau=0.11,0.12,0.13,14,15,..,0.19$ respectively. } 
		\label{fig3}
		\end{figure} 

        In this section, we wish to study the entanglement of the dynamical states after a finite round of measurement. 
Here, we divide the system into $L/4:3L/4$ bi-partitions and calculate the entanglement between the two parts, considering the total system size $L=18,20,22,24,26$. It is to be noted that, for the system size $L\neq4n$, where $n\in \mathbb{Z}$, we calculate the average entanglement of the two bi-partition $(L-2)/4:3(L-2)/4$ and $(L+2)/4:3(L+2)/4$. In the inset of panel (a) of Fig.~\ref{fig3}, we plot the entanglement entropy with system size for TFIM with longitudinal field. Each curve corresponds to a different $\tau$ value. Here the parameters of the considered Hamiltonian is chosen as $h_z=0.5$ and $h_x=0.6$. The $\tau$ values are considered to be $\tau=0.09,0.10,0.11,....,0.16$. We find that the entanglement remains almost constant with the system size upto $\tau=0.13$ and for $\tau>0.13$ the entanglement increases linearly with system size. This shows an area-to-volume law transition about the tau value $\tau_c=0.13$, around the point where the convex-to-concave transition is captured by survival probability. Further to have a closer look in this area-to-volume law transition, another diagnostic that is widely used for detecting the area-to-volume law transition around the transition point is captured by the scaling relation,

\begin{equation}\label{eq.9}
S(\tau)-S(\tau_c)
=
F\!\left((\tau-\tau_c)L^{1/\nu}\right).
\end{equation}

This equation indicates that if a phase transition exists in entanglement, the $S$ vs $L$ curves for different system sizes $L$ will collapse onto each other around the critical point $\tau_c$ and for appropriate value of $\nu$.
In the main plot of Fig.~\ref{fig3}, we plot $S(L/4,\tau)-S(L/4,\tau_c)$ vs. $(\tau-\tau_c)L^{1/\nu}$ plot and use a numerical optimization procedure as explained  in  Appendix~\ref{App_B} to find the best collapse over a range of $\tau_c$ and $\nu$. We found a proper scaling collapse for $\tau=0.127$ and $\nu=1.68$, showing a clear transition of area to volume law transition around $\tau_c \approx 0.13$. The similar analysis for ANNNI model with $\kappa=0.2$ and $h_z=-0.2$ is shown in panel (b) of Fig.~\ref{fig3}. From the inset in panel (b) in Fig.~\ref{fig3}, in $S(L/4,\tau)$ vs. $L$ graph, a rough estimate of the critical point is observed at $\tau \geq0.17$ from where the entanglement starts increasing with system size and for $\tau<0.17$ the entanglement almost remains constant with system size. Further, from the scaling analysis using Eq.~\eqref{fig3}, we find a proper collapse around $\tau\approx0.168$ with $\nu=1.247$ indicating the transition point $\tau_c\approx0.168$. The convex-to-concave transition point, that is captured by the survival probability is also lies around the same point $\tau_c=0.17$.

The above analysis shows that an area-to-volume law transition is present in both the TFIM with longitudinal field and the ANNNI model. We also note that the $\tau_c$ extracted from the survival probability is the same as the $\tau_c$ of the entanglement transition. This establishes that the survival probability peaks are signatures of the entanglement transitions occurring due to measurements.
\section{Initially entangled GHZ state.}\label{IV}

In this section, we deal with a different initial state than the transverse-polarized state $\ket{0}^{\otimes L}$. To this end we consider GHZ state as the initial state, which can be achieved by putting $\phi=\pi/4$ in Eq.~\eqref{eq6}.

\subsection{Detection of transition point by survival probability for initial GHZ state} \label{IVA}

\begin{figure*}
		\centering
			\includegraphics[width=17.5cm]{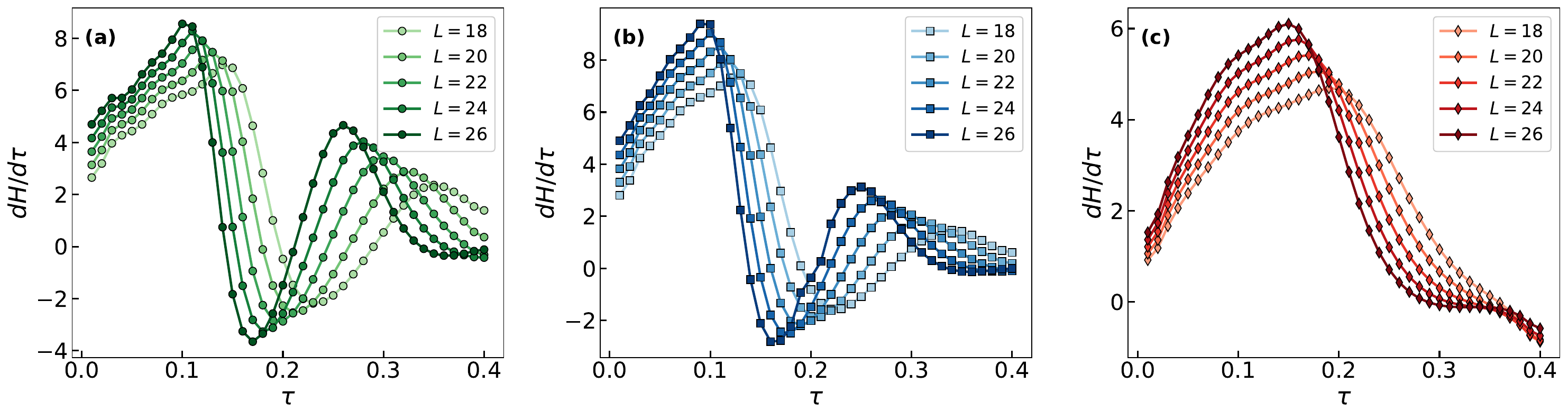}
\caption{\textbf{Variation of survival probability for initial GHZ state.}
Panel (a) shows the behavior of $\frac{dH}{d\tau}$ with $\tau$ for different system sizes $L=18,20,22,24,26$ for the transverse field Ising model in absence of any longitudinal field, viz. $h_x=0$ and $h_z=0.5$. Panel (b) show the same $\frac{dH}{d\tau}$ vs $\tau$ plot for TFIM with nonzero longitudinal magnetic field $h_x=0.6$ and $h_z=0.5$ for the same system sizes considered in panel (a). In panel (c) also we have done same analysis in ANNNI model with $\kappa=0.2$ and $h_z=0.2$ for the system sizes exactly considered in panel (a) and (b).  } 
		\label{fig4}
		\end{figure*} 

 \begin{figure}
		\centering
			\includegraphics[width=8.5cm]{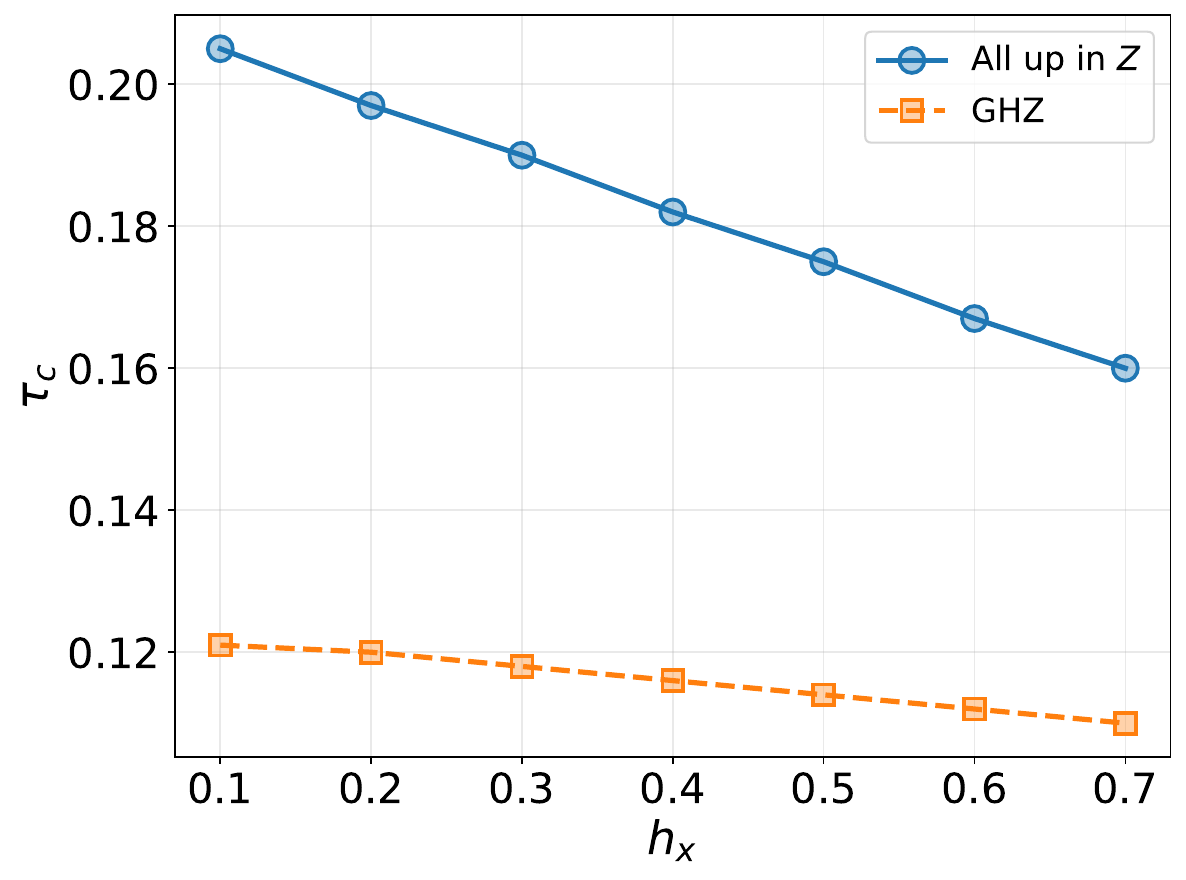}
\caption{\textbf{Dependence of transition point with longitudinal field strength.}
In this figure, the horizontal axis denotes the longitudinal field strength ($h_x$) and the vertical axis denotes the transition point ($\tau_c$). The curve corresponding to initially $+Z$ polarized configuration is denoted by blue line with circular marker and the curve corresponding to GHZ state is denoted by red curve with square markers.   } 
		\label{fig44}
		\end{figure}

    \begin{figure*}
    \centering
    \includegraphics[
        width=17cm,height=4.8cm,
    ]{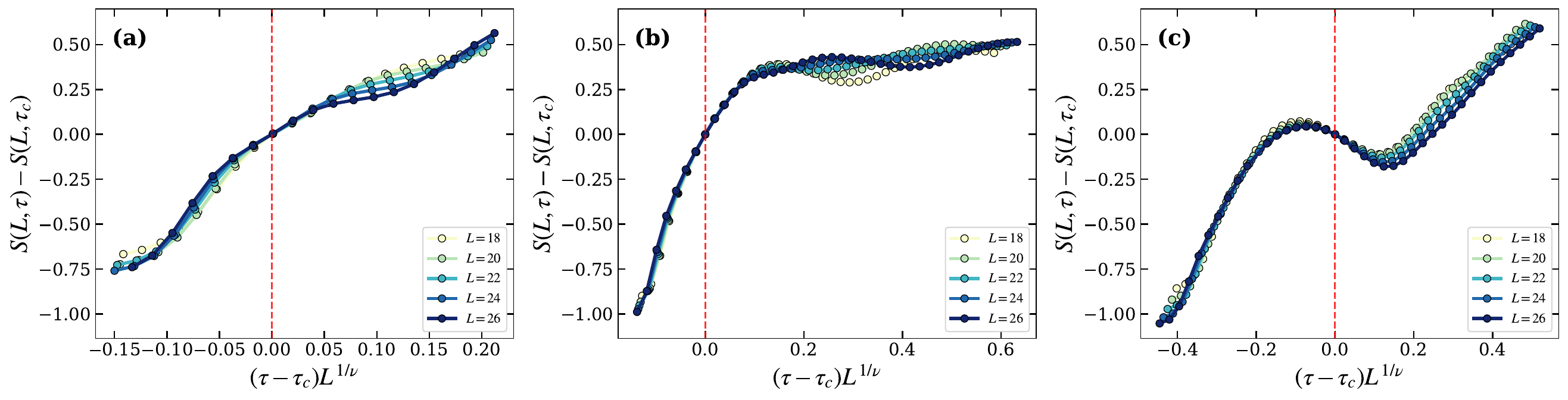}
    \caption{\textbf{Scaling collapse of entanglement entropy.} Scaling collapse of the entanglement entropy, shown by plotting $S(L,\tau)-S(L,\tau_c)$ as a function of the rescaled variable $(\tau-\tau_c)L^{1/\nu}$. Panel (a) presents the results for the TFIM with $h_z=0.5$. Panel (b) corresponds to the TFIM with longitudinal field with $h_z=0.5$ and $h_x=0.6$. Panel (c) shows the results for the ANNNI with $h_z=0.2$ and $\kappa=0.2$.}
   
    \label{GHZ_sceal}
\end{figure*}
Here, we study the nature of survival probability in our measurement-based protocol where the initial state is GHZ state.

 In Fig.~\ref{fig4}, Panel (a), we consider  TFIM without the effect of any longitudinal field and focus on the behavior of the derivative of the height of the plateau $\frac{dH}{d\tau}$ with respect to $\tau$, where the initial entangled GHZ state for the system sizes $L=18,20,22,24,26$. The plot shows that $\frac{dH}{d\tau}$ has two local maximum which was not present in case of the initial state $\ket{0}^{\otimes L}$.
Now here comes a natural question: which peak actually indicates a phase transition point? The  answer to this question is given in the Sec.~\ref{IVB}. Similar study is done in panel (b) for TFIM with longitudinal field
with longitudinal field strength $h_x=0.6$ and transverse field $h_z=0.5$. Here we plot $\frac{dH}{d\tau}$ vs. $\tau$ plot for the same system sizes considered in panel (a). A similar feature like integrable TFIM is captured here showing two peaks almost in the same $\tau$ values of corresponding system sizes as that of TFIM that shown in panel (a) with a difference that the heights of the second peaks corresponding to TFIM with longitudinal field are smaller in comparison to integrable TFIM. For both TFIM and TFIM with longitudinal field, the first peak arises nearly $\tau_c\approx 0.1$ In panel (c) we consider ANNNI model and plot $\frac{dH}{d\tau}$ vs. $\tau$ plot for the parameters $\kappa=0.2$ and $h_z=-0.2$ and considering same system sizes. In contrast to integrable and TFIM with longitudinal field, we see that there exists only one prominent peak in the $\frac{dH}{d\tau}$ vs. $\tau$ around $\tau_c\approx 0.17$.

At the same time we observe how the transition point behaves with the longitudinal field $h_x$ for both the initially $+Z$ polarized state and GHZ state. In Fig.~\ref{fig44}, show the $\tau_c$ vs $h_x$ curve for $+Z$ polarized state and GHZ state denoted by blue smooth curve with circular dots and red dashed curve with square dots.
For both type of initial state the transition point $\tau_c$ decreases and goes to zero as we increases $h_x$. Here the most intriguing point is that the $\tau_c$ goes more faster towards zero for initially $+Z$ polarized state then GHZ state. It implies that for GHZ state the transition persists for a  wider range of the integrability-breaking $h_x$  field values than the initially $+Z$ polarized state.

\subsection{Detection of transition point by entanglement for GHZ state}\label{IVB}

Further, we also see the collapse of entanglement entropy curves according to the scaling function given in Eq.~\eqref{eq.9} for three types of models, i.e TFIM, TFIM with longitudinal field and ANNNI model where the initial state is chosen to be the GHZ state. For TFIM, we consider $h_z=0.5$.  We consider the system sizes $L=18,20,22,24,26$. The analysis is shown in panel (a) of Fig.~\ref{GHZ_sceal}. We obtain the scaling collapse is exactly occurring $\tau=0.108$ and for $\nu=4.31$. Similarly in panel (b) we show the scaling collapse for TFIM with field strength $h_z=0.5$ and $h_z=-0.6$ respectively for the same considered system sizes as in panel (a). The collapse is obtained for $\tau=0.105$ and $\nu=3.69$. In panel (C) we do the same analysis for the ANNNI model and show that the collapsing is occuring arround $\tau=0.19$ with $\nu=4.71$. The scaling collapse for all three models are occuring about the same point where $\frac{dH}{d\tau}$ vs. $\tau$ curve shows peak in Fig.~\ref{fig4}. Most important point to note that although the there are two peaks arising in $\frac{dH}{d\tau}$ vs. $\tau$ in Fig.~\ref{fig4} (a) and (b) for TFIM and TFIM with longitudinal field, the transition captured by entanglement is around the first peak about the point $\tau_c\approx0.1$.

\section{Scaling analysis
}\label{V}
In this section, we analyze how the transition points extracted from the survival probability scales with systems sizes. For the TFIM, Ref.~\cite{Suv_sp} shows that the survival probability can be calculated recursively, starting from a configuration in which all the spins are initially polarized in $+Z$ direction. It should be noted that bipartite entanglement entropy cannot be calculated analytically or recursively at large system sizes. So the survival probability as a detector of the entanglement transition becomes important while we are extending the study to large system sizes.

We have shown in Ref.~\cite{Suv_sp} that 
the state $\ket{\psi_{n}}$ after $n$th measurement can be written as,
 \begin{equation}
     \ket{\psi_{n}}=\sum_{m=0}^n C_m^{(n)}\ket{\phi_m}.
 \end{equation}
 where $\ket{\phi_n}=e^{-iH\tau n}\ket{I}$ and the coefficients $C_m^{(n)}$ follows the 
recursion relations
\begin{equation} C^{(n+1)}_0=-\sum_{m=0}^n C^{(n)}_mf_{m+1}, \;\;\;  C^{(n+1)}_m=C^{(n)}_{m-1}  \label{recursion1} \end{equation}
for $0<m\le n$. This enables one to compute the survival probability at the $n$-th step,
\begin{equation}
R_n
=
\langle\psi_n|\psi_n\rangle
=
\sum_{m_1,m_2=0}^{n}
\left(C_{m_1}^{(n)}\right)^*
C_{m_2}^{(n)}
f_{m_2-m_1}
\label{eq:20}
\end{equation}
provided that the quantity $f_n$ is known for all the $n$ steps. 

 For TFIM Hamiltonian, the quantity $f_n$ has been calculated for $\ket{0}^{\otimes L}$ initial state as given in Ref.~\cite{Suv_sp}. The $f_n$ corresponding to generalized GHZ state for TFIM is given by
\begin{equation}
\begin{aligned}
f_n &=  \cos^2 \phi \prod_k \left[ \cos(\lambda_k t)
+ i \sin(\lambda_k t)\cos(2\theta_k) \right] \\
&\quad + \sin^2 \phi \prod_k \left[ \cos(\lambda_k t)
- i \sin(\lambda_k t)\cos(2\theta_k) \right] \\
&\quad + \sin \phi \cos \phi \left(1 + (-1)^{N/2}\right)
\prod_k \left[ \sin(2\theta_k)\sin(\lambda_k t) \right].
\end{aligned}
\end{equation}
where  $\exp(2i\theta_k)=(\Gamma + \exp(i k))/\lambda_k$ and
$\lambda_k = 2\sqrt{\Gamma^2 + 1 + 2\Gamma \cos k}$, $k=(2n+1)\pi/L$ with $n=0,1,2,\cdots,L/2-1$. The detail calculation is given in appendix~\ref{App_CB}

Although we are able to calculate the $f_n$ analytically for integrable TFIM model, for the TFIM with longitudinal field and ANNNI, we are unable to calculate the $f_n$ analytically. To bypass that, we have used the fact that for a given, small $n$ and $\tau$ the quantity $f_n^{1/L}$
remains independent of  $L$ for system sizes upto $L=26$, i.e,
\begin{equation}\label{eq117}
f_n^{1/L}=\text{constant}.
\end{equation}

 If we assume that this same property will hold for larger system sizes, we can calculate the survival probability $R_n$ for larger system sizes for small $n$ and $\tau$, using this property in Eq.~\eqref{eq117}. As the transition points shifts towards small tau value and the first plateau extends and becomes more flat as system size increases, we are able to detect the transition point for larger system sizes for non-integrable TFIM with longitudinal field and ANNNI model if we use a small $n$ value for the calculation of $R_n$. 

\subsection{Scaling analysis for initially $+Z$ polarized state
}\label{VA}

\begin{figure}
		\centering
			\includegraphics[width=8.5cm]{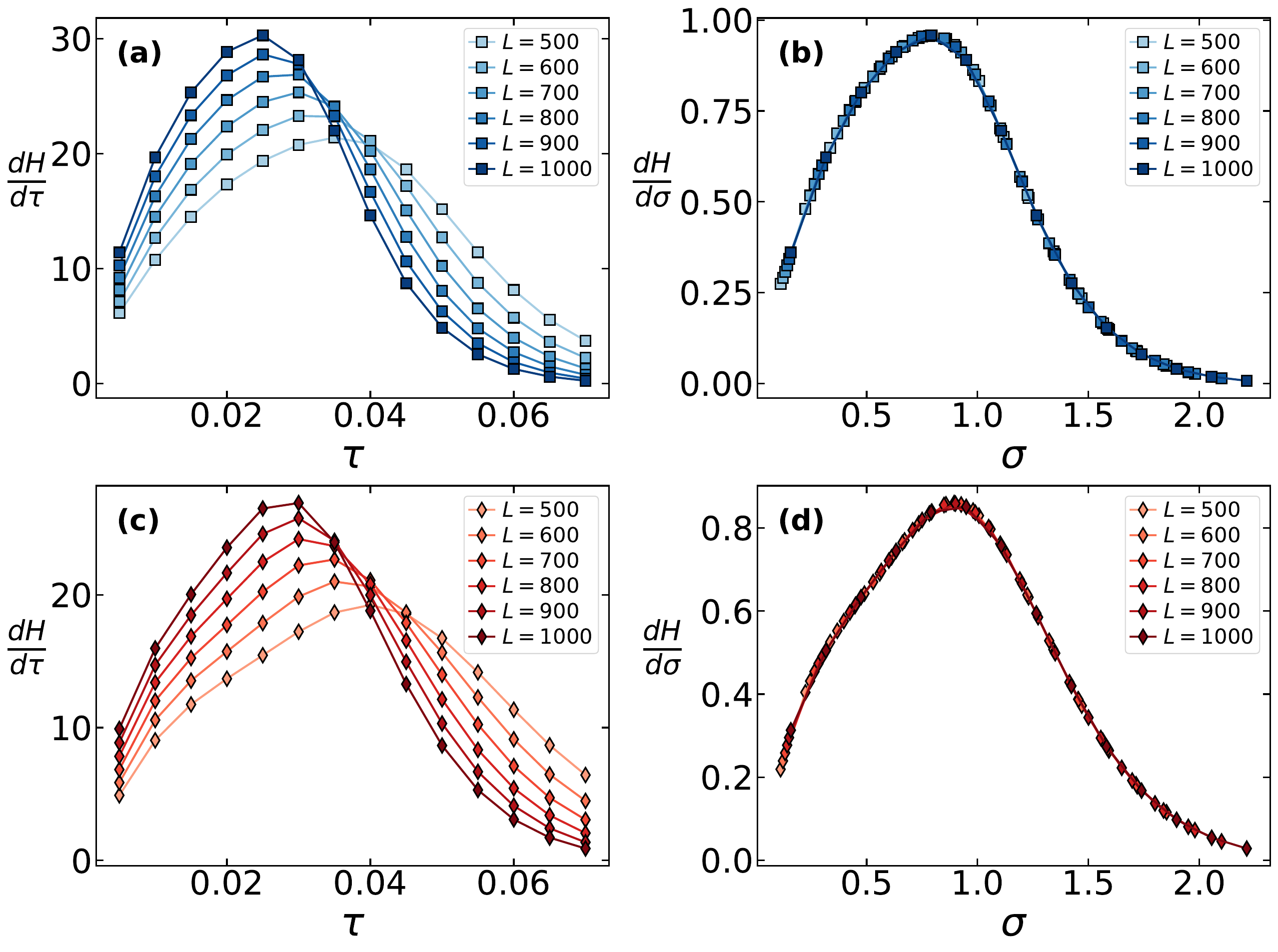}
\caption{\textbf{Dependence of the plateau height with the interaction period for larger system sizes for initially $+Z$ polarized state.}
Panel (a) shows the variation of $\frac{dH}{d\tau}$ as a function of $\tau$ for different system sizes in the transverse-field Ising model with an additional longitudinal field, with parameters $h_z=0.5$ and $h_x=0.6$. Different system sizes are represented by sequential blue shades with square markers. Panel (b) presents the corresponding finite-size scaling analysis by plotting $\frac{dH}{d\sigma}$ as a function of $\sigma=\tau\sqrt{L}$ for the same model and parameter set. Panels (c) and (d) display the analogous results for the ANNNI model, corresponding respectively to the analyses shown in panels (a) and (b). For the ANNNI model, the parameters are chosen as $\kappa=0.2$ and $h_z=0.2$. In all panels, the system sizes considered are $L=500,600,700,800,900$, and $1000$. } 
		\label{fig6}
		\end{figure}

Here, we perform the scaling analysis of the transition point $\tau_c$ for TFIM with longitudinal field and ANNNI models. As evident from Fig.~\ref{fig6}, the transition point shifts towards smaller values of $\tau$ with increasing system size. Here, we consider the system sizes $L=500, 600, 700, 800, 900,$ and $1000$ and the $n$ value for all system sizes is $n=4$ is justified because it lies within first plateau. In panel (a) of Fig.~\ref{fig6}, we plot the derivative of the plateau height of the survival probability with respect to the measurement interval $\tau$, namely $\frac{dH}{d\tau}$, as a function of $\tau$ for different system sizes for TFIM with longitudinal field at $h_z=0.5$ and $h_x=-0.6$. Curves corresponding to different system sizes are represented by sequential shades of blue. As the system size increases, the transition point systematically shifts towards smaller values of $\tau$, indicating its tendency to approach zero in the thermodynamic limit. 

A similar analysis is performed for the ANNNI model in Fig.~\ref{fig6} (c), where the results for different system sizes are shown using sequential shades of red for $h_z=-0.2$ and  $\kappa=0.2$. Once again, the transition point is observed to move towards $\tau=0$ with increasing system size, suggesting the disappearance of the transition in the thermodynamic limit.

To further investigate the finite-size scaling behavior, we rescale the horizontal axis by introducing the scaling variable $\sigma=\tau\sqrt{L}$ and plot $\frac{dH}{d\sigma}$ as a function of $\sigma$ for both the mixed-field Ising and ANNNI models. As shown in Fig.~\ref{fig6} (b) and (d), the data corresponding to different system sizes exhibit an excellent collapse onto a single universal curve. This scaling collapse implies that the finite-size critical point obeys the relation
\begin{equation}\label{eq18}
\tau_c^{L}=\frac{\tilde{C}}{\sqrt{L}},
\end{equation}
where $\tau_c^{L}$ denotes the transition point for a system of size $L$, and $\tilde{C}$ is a model-dependent constant.

Notably, the same scaling behavior has previously been reported for the integrable transverse-field Ising model in the absence of a longitudinal field. However, the inclusion of a longitudinal field reduces the proportionality constant $\tilde{C}$, causing the transition point to approach zero more rapidly with increasing system size and thereby accelerating the disappearance of the phase transition. This behavior of $\tilde{C}$ with $h_x$ is described in Appendix~.\ref{App_BB}.

\subsection{Scaling analysis for initially GHZ state
}\label{VB}

In this section we do the scaling analysis of 
for the initial GHZ state for the system sizes $L=70,80,90,.....,150$ for all three considered models and fit the curves with the scaling function

\begin{equation}
    \tau_c=\frac{\bar{C}_{\mathcal{G}}}{L^{\alpha}}.
\end{equation}
In this study for all three systems, for fair comparison we consider the system size the field strength $h_z=0.5$ for all the models and $h_x=0.6$ for TFIM with longitudinal field $\kappa=0.2$ for ANNNI.

\begin{figure}
		\centering
			\includegraphics[width=8.6cm,height=6.4cm]{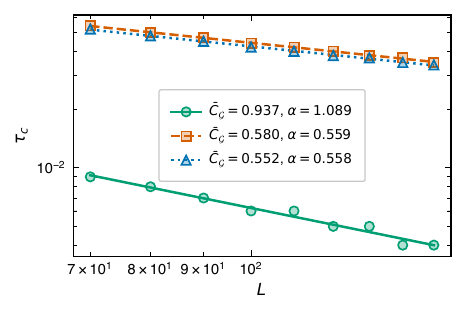}
\caption{\textbf{Scaling analysis for the initial GHZ state for three considered model} Here the horizontal and the vertical axes denote the system size $L$ and the transition point $\tau_c$ respectively. The axes are displayed in log scale. The green, blue and red dots correspond to TFIM without longitudinal field, TFIM with longitudinal field and ANNNI model and the lines with the same colour  as that of the dots corresponding to the fitting curve.
} 
		\label{GHZ_scalee}
		\end{figure}

Fig.~\ref{GHZ_scalee} presents the finite-size scaling analysis for the initially GHZ state in the three models considered. Both the horizontal and vertical axes are displayed on logarithmic scales, representing the system size (L) and the  $\tau_c$, respectively. The green, blue, and red dots correspond to the integrable TFIM, the TFIM with a longitudinal field, and the ANNNI model, respectively. The solid lines of the corresponding colors represent the best-fit power-law curves for the respective datasets.

\begin{table}[h]
\centering

\label{table.1}
\begin{tabular}{|l|c|c|}
\hline
\textbf{Model} & ${\bar{C}_{\mathcal{G}}}$ & $\mathbf{\alpha}$ \\
\hline
TFIM     & 0.937 & 1.089 \\
\hline
Mixed-field TFIM   & 0.552 & 0.558 \\
\hline
ANNNI model         & 0.580 & 0.559 \\
\hline

\end{tabular}
\caption{Fitted values of the average generalized geometric measure,
$\bar{C}_{\mathcal{G}}$, and the scaling exponent $\alpha$ for the three
models considered in this work.}
\end{table}

From Fig.~\ref{GHZ_scalee}, one can note that the transition point corresponds to all three models going to zero at thermodynamic limit but the interesting point to note that when we introduce integrability breaking terms like longitudinal field and next-to-next-nearest interaction, the transition point moves slowly towards a zero value with system size $L$, implying that the transition exists even at higher system sizes than the integrable TFIM. This feature is also evident from table~\ref{table.1}. The value of $\alpha$ for the integrable TFIM is significantly larger than those for the other two models, indicating a faster decay of $\tau_c$ with increasing system size.

    \section{conclusion}\label{VI}
In this paper, we have studied the effect of non-integrability and of initial state entanglement on the measurement induced phase transition (MIPT) occurring due to repeated global measurements. We have started from an initial state and unitarily evolved it under transverse field Ising model with and without longitudinal field and transverse ANNNI model and periodically measured the state in a basis with two elements, the initial state projector and the projector orthogonal to it. After each measurement, we select the state in the orthogonal subspace of the initial state. Though the MIPT point moves towards zero with increasing system size under this protocol, the critical point moves relatively slowly with system size in non-integrable dynamics when the initial state is entangled. which implies the phase transition exists in a relatively larger system for the non-integrable model than the integrable one for the initially entangled state. When the initial state is a product state, the non-integrability is not able to make the scaling slower (with respect to system size) over integrable dynamics.

Summarizing the important results of this paper, we have first established that the survival probability detects the transition point where the bipartite entanglement entropy shows a transition from area law to volume law scaling. This is being done in small system sizes upto $L=26$. Assuming that survival probability is a good diagnostic for MIPT, the survival probability is calculated at large system sizes. At large system sizes, MIPT point $\tau_c$ scales as $1/\sqrt{L}$ when the initial state is a product state, both for integrable and non-integrable dynamics. But when the initial state is entangled, the MIPT points scales approximately as $1/L$ for integrable dynamics and as $1/L^{0.55}$ for non-integrable dynamics. The scaling becomes apparent at system sizes over $L=100$.

Hence we have established that MIPT under global measurement persists in both integrable and non-integrable dynamics at finite system sizes. Additionally, we have shown that non-integrability slows down the approach of $\tau_c$ towards zero when the initial state is entangled. 

\acknowledgements 
  PC acknowledges support from ‘INFOSYS scholarship for senior students’ at Harish Chandra Research Institute, India. PN acknowledges Harish Chandra Research Institute for access to their infrastructure. US acknowledges financial support from the Anusandhan National Research Foundation (ANRF), Government of India, under the Grant No. ANRF/ARG/2025/004617/PS.
    \twocolumngrid
\section*{Appendix}

\label{sec-Appendix}
\appendix

\appendix

\section{System-size independent behaviour of $f^{1/L}_n$}
\begin{figure}[h]
		\centering
			\includegraphics[width=8.5cm]{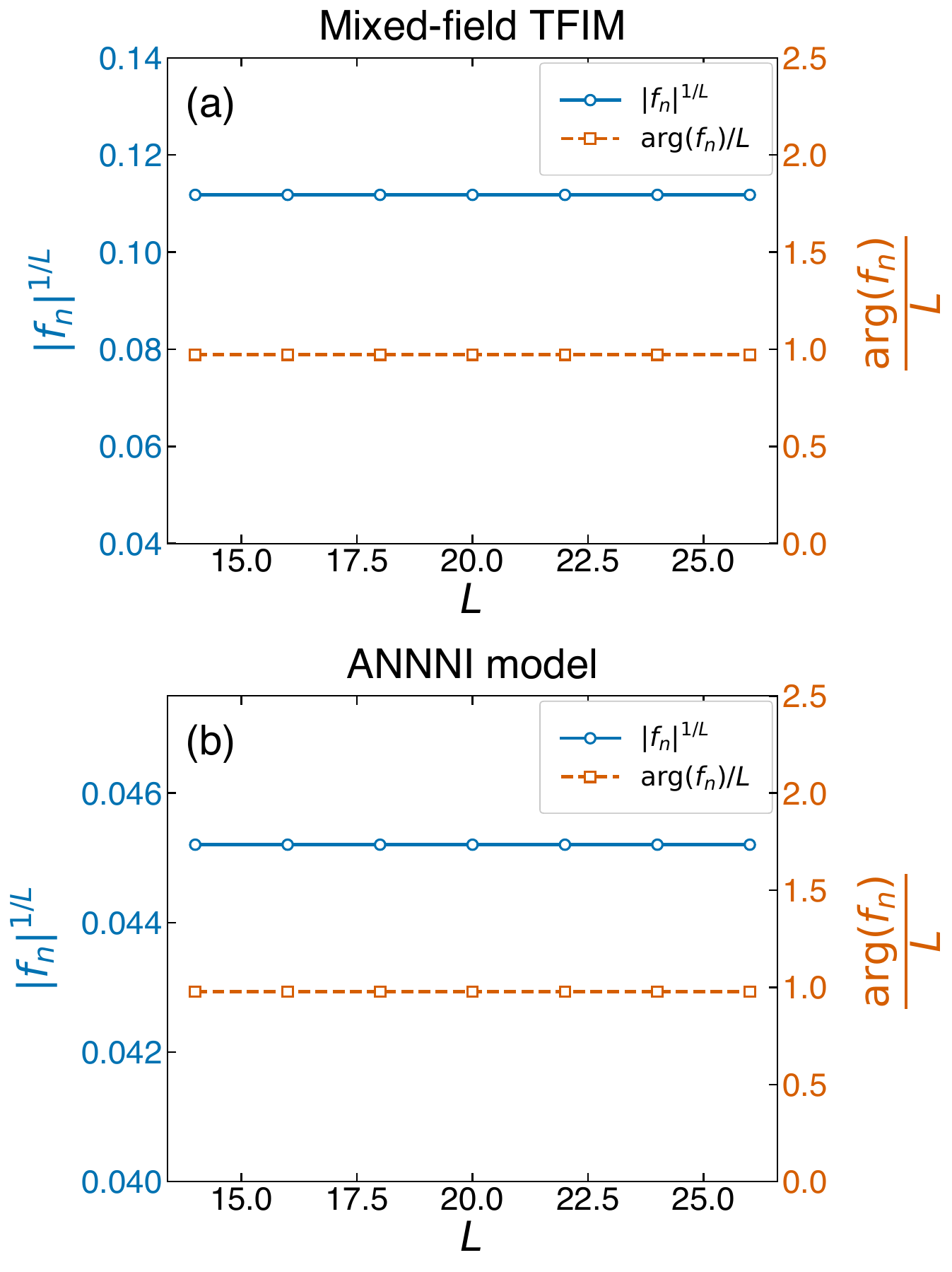}
\caption{\textbf{Nature of $|f_n|^{1/L}$ and $\arg (f_n)/ L$ with system size L}
Panel~(a) shows the plots of $|f_n|^{1/L}$ and $\arg(f_n)/L$ as functions of $L$ for the mixed-field TFIM with $h_x=0.6$ and $h_z=0.5$. Panel~(b) shows the corresponding plots for the ANNNI model with $\kappa=0.2$ and $h_z=0.2$. In both panels, the evolution time is fixed at $\tau=0.07$.} 
		\label{fig8}
		\end{figure} 

Although the quantity $f_n=\bra{I}\exp(-i\mathcal{H}n\tau)\ket{I}$ can be evaluated analytically for integrable models such as the TFIM, an analytical calculation is generally not possible for non-integrable systems, including the ANNNI model and the TFIM with a longitudinal field, particularly for large system sizes. Since $f_n$ is, in general, a complex quantity, it can be expressed as $f_n=|f_n|\exp(-i\alpha)$, where $\alpha=\arg(f_n)$. Consequently, $f_n^{1/L}=|f_n|^{1/L}\exp(-i\alpha/L)$. Fig.~\ref{fig8} separately illustrates the dependence of $|f_n|^{1/L}$ and $\alpha/L=\arg(f_n)/L$ on the system size $L$. It is evident that both quantities remain essentially independent of $L$, indicating their convergence to well-defined values in the thermodynamic limit. Panels~(a) and (b) correspond to the TFIM with longitudinalfield and the ANNNI model, respectively. For the TFIM with longitudinal field, the parameters are chosen as $h_x=0.6$ and $h_z=0.5$, while for the ANNNI model, they are $\kappa=0.2$ and $h_z=-0.2$. In both models, the evolution time is fixed at $\tau=0.70$.

\section{Derivation of the $f_n$ for initially generalized GHZ state}\label{App_CB}

The generalized GHZ state can be written as,
\begin{equation}
|\psi_0\rangle^{\mathcal{G}}_{GHZ} = \cos \phi\, |U\rangle + \sin \phi\, |D\rangle,
\end{equation}

where $|U\rangle := \ket{0}^{\otimes L}, 
\hspace{0.2cm}$ and 
$|D\rangle := \ket{1}^{\otimes L}$. For $\phi=\pi/4$ the state become maximally entangled GHZ state.
Now starting from the initial state $\ket{\psi_0}^{\mathcal{G}}_{GHZ}$, the amplitude corresponding to the initial state, after the evolution for a time $t=n\tau$, will be
\begin{equation}
f_n = \langle \psi_0 | e^{-i H n \tau} | \psi_0 \rangle.
\end{equation}
This quantity $f_n$ can be evaluated for integrable TFIM using Jordan-Wigner transformation \cite{JD}. 
Under the Jordan-Wigner transformation , the states $|U\rangle$ and $|D\rangle$ correspond to
\begin{equation}
|U\rangle = \prod_k |11\rangle_k, 
\qquad 
|D\rangle = \prod_k |00\rangle_k.\nonumber
\end{equation}

Since the transverse field Ising model Hamiltonian decomposes as $H = \sum_k H_k$, where $H_k$ acts on the independent momentum sector corresponding to $(k,-k)$ momentum pair is given by

\begin{eqnarray}\nonumber
\mathcal{H}_k
=&
-&2i\sin k
\left[
a_k^\dagger a_{-k}^\dagger
+
a_k a_{-k}
\right]\\\nonumber
&-&
2(h_z+\cos k)
\left[
a_k^\dagger a_k
+
a_{-k}^\dagger a_{-k}
-1
\right]\nonumber,
\end{eqnarray}
where   $k=(2n+1)\pi/L$ with $n=0,1,2,\cdots,L/2-1$.
Therefore, time evolution operator will be factorized as
\begin{equation}
e^{-i H t} = \prod_k e^{-i H_k t}, \qquad t = n\tau.\nonumber
\end{equation}

If we apply the unitary operator $\prod e^{-i H_k t}$ on the initial state $\ket{\psi_0}_{GHZ}$, it acts independently on each $\ket{00}_k$ and $\ket{11}_k$.
Therefore after the action of unitary the terms $\ket{00}_k$ and $\ket{11}_k$ will be
\begin{align}
e^{-i H_k t} |11\rangle_k &=
\Big[ \cos(\lambda_k t)
+ i \sin(\lambda_k t)\cos(2\theta_k) \Big] |11\rangle_k \nonumber\\
&\quad
- \sin(2\theta_k)\sin(\lambda_k t)\, |00\rangle_k, \\
e^{-i H_k t} |00\rangle_k &=
\sin(2\theta_k)\sin(\lambda_k t)\, |11\rangle_k \nonumber\\
&\quad
+ \Big[ \cos(\lambda_k t)
- i \sin(\lambda_k t)\cos(2\theta_k) \Big] |00\rangle_k .
\end{align}
where,
 $\cos(2\theta_k)=(\Gamma + \cos k)/\lambda_k$ and $\sin(2\theta_k)= \sin k/\lambda_k$ and the $\lambda_k$ is given by
$
\lambda_k = 2\sqrt{\Gamma^2 + 1 + 2\Gamma \cos k}
$. The term $f_n$ will contain four terms as written below
\begin{equation}\label{Eq_14}
\begin{aligned}
F_{UU}(t) &= \langle U| e^{-i H t} |U\rangle\\
&= \prod_k \Big[\cos(\lambda_k t)
+ i \sin(\lambda_k t)\cos(2\theta_k)\Big], \\
F_{DD}(t) &= \langle D| e^{-i H t} |D\rangle\\
&= \prod_k \Big[\cos(\lambda_k t)
- i \sin(\lambda_k t)\cos(2\theta_k)\Big], \\
F_{UD}(t) &= \langle U| e^{-i H t} |D\rangle\\
&= \prod_k \Big[\sin(2\theta_k)\sin(\lambda_k t)\Big], \\
F_{DU}(t) &= \langle D| e^{-i H t} |U\rangle\\
&= (-1)^{N/2}\prod_k \Big[\sin(2\theta_k)\sin(\lambda_k t)\Big].
\end{aligned}
\end{equation}


Now, using the expression of $F_{UU}$, $F_{DD}$, $F_{UD}$ and $F_{DU}$ provided in Eq.~\eqref{Eq_14} $f_n$ can be written as

\begin{equation}
\begin{aligned}
f_n &= \langle \psi_0 | e^{-iHt} | \psi_0 \rangle \\
&= \cos^2 \phi\, F_{UU}(t) + \sin^2 \phi\, F_{DD}(t)\\
&+ \sin \phi \cos \phi \left[ F_{UD}(t) + F_{DU}(t) \right] \\
&= \cos^2 \phi \prod_k \left[ \cos(\lambda_k t)
+ i \sin(\lambda_k t)\cos(2\theta_k) \right] \\
&\quad + \sin^2 \phi \prod_k \left[ \cos(\lambda_k t)
- i \sin(\lambda_k t)\cos(2\theta_k) \right] \\
&\quad + \sin \phi \cos \phi \left(1 + (-1)^{N/2}\right)
\prod_k \left[ \sin(2\theta_k)\sin(\lambda_k t) \right].
\end{aligned}
\end{equation}

\section{Variation of $\tilde{C}$ with $h_x$}\label{App_BB}

\begin{figure}[h]
		\centering
			\includegraphics[width=8.5cm]{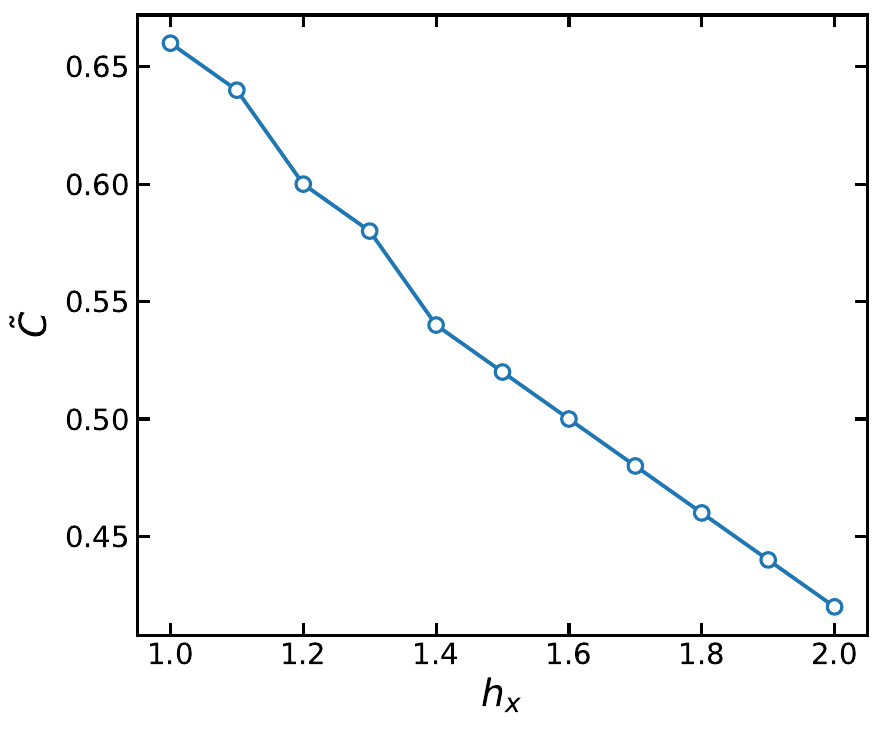}
\caption{\textbf{Variation of $\tilde{C}$ with $h_x$ for $h_z=0.5$}
} 
		\label{fig9}
		\end{figure}

In Sec.~\ref{VA}, we have shown that the transition point $\tau_c$ scales as $1/\sqrt{L}$ for both the non-integrable models similar to TFIM. Although the proportionality constant $\tilde{C}$ in Eq.~\eqref{eq18} varies with the strength of the integrability-breaking term. Here in Fig.~\ref{fig9}, we have shown how $\tilde{C}$ varies with the longitudinal field strength $h_x$ for mixed-field TFIM.

We have shown that as $h_x$ increases the proportional constant decreases. The whole study is done for $h_z=0.5$.

\section{Numerical method of estimating the exponent $\nu$}\label{App_B}
The critical exponent $\nu$ is extracted through a finite-size data-collapse analysis. We first generate $n$ data sets for $S$ as a function of $\tau$, each corresponding to a different system size $L$. For each trial pair $(\tau_c,\nu)$, the data are rescaled according to
\begin{equation}
S(\tau)-S(\tau_c)
\quad \text{versus} \quad
(\tau-\tau_c)L^{1/\nu}.
\end{equation}

Here, $S(\tau_c)$ is evaluated separately for each system size. Whenever $\tau_c$ does not coincide with one of the sampled values of $\tau$, the corresponding value of $S(\tau_c)$ is estimated using linear interpolation.

Next, we consider a symmetric interval of width $2\tau_0$ centered at the trial critical point, i.e., $[\tau_c-\tau_0,\tau_c+\tau_0]$, and partition it into several bins determined by the resolution of the raw data. Within each bin, the rescaled values $S(\tau)-S(\tau_c)$ obtained for different system sizes are grouped together. The mean of these values is then computed, and the corresponding least-squares deviation from the bin average is evaluated. The total least-squares error is defined as the sum of these deviations over all bins. The optimal estimates of $\tau_c$ and $\nu$ are subsequently identified as the values for which the total least-squares error is minimized.

\bibliography{MIPT}

\end{document}